\documentclass{article}

\usepackage{amsmath}
\usepackage{amssymb}
\usepackage{amsfonts}
\usepackage{graphicx}
\usepackage{latexsym}

\begin{document}

\title{Progress in the mathematical theory of quantum disordered systems}
\author{Walter F. Wreszinski\\
Instituto de F\'{i}sica\\
Universidade de S\~{a}o Paulo\\
Caixa Postal 66318\\
05314-970 S\~{a}o Paulo, SP , Brazil}

\maketitle

\begin{abstract}
We review recent progress in the mathematical theory of quantum
disordered systems: the Anderson transition (joint work with Domingos
Marchetti), the (quantum and classical) Edwards-anderson spin-glass
model and return to equilibrium for a class of spin glass models,
which includes the EA model initially in a very large transverse
magnetic field. 
\end{abstract}

\textbf{1 - Introduction}

In recent years there has been a significant progress in the
mathematical theory of (quantum) disordered systems. Our purpose here
is to present the main ideas (without proofs), with a clear discussion
of their conceptual and physical relevance, as well as a brief
comparison with the recent, analogous results in the literature. Some
new remarks and results, which clarify some important conceptual
points, are also included. 

We shall be interested in properties of disordered systems at low
temperatures, i.e., near the ground state (absolute temperature $T =
0$); in the critical region of spin glasses, there exist the
spectacular recent rigorous results on the mean field theory (see the
review by F. Guerra \cite{Gue}), and various results for other
disordered systems, including the well understood high temperature
phase in spin glasses, are discussed in the comprehensive book by
A. Bovier \cite{Bo}. 

Our restriction to very low temperatures implies, of course, that we
shall be dealing exclusively with quantum systems. In particular, the
Ising model, when it appears, must be regarded as the anisotropic
limit of quantum (spin) systems: the fact that this is not only
mathematically so is emphatically demonstrated by the well-known fact
that the critical exponents of the Ising model in three dimensions
(see, e.g., \cite{Zinn}) are surprisingly close to those measured in
real magnetic systems, precisely because most of the latter are highly
anisotropic.  

Three important issues appearing in the above-mentioned context are
\textbf{localization}, arising in connection with the Anderson
transition, which we discuss in section 2, reporting on joint work
with Domingos Marchetti (\cite{MW},\cite{MWB}), \textbf{frustration},
which appears in short-range spin glasses such as the Edwards-Anderson
(EA) model, discussed in section 3, and the (thermodynamic version of)
\textbf{instability of the first and second kind}, which relates to
the return to equilibrium for special initial states and probability
distributions in a class of models (including the EA model) in section
4.

\textbf{2 - The Anderson transition}

The breakdown of translation invariance (or, more generally, Galilean
invariance in many-body systems with short-range forces) leads to the
existence of \textbf{crystals}, and the corresponding Goldstone
excitations are phonons \cite{Swi}. On the other hand, the discrete
translation invariance subgroup of the crystal is also frequently
broken by e.g. impurities. This fact brings about a number of
important new conceptual issues, and is usually modelled by the
introduction of a random local potential in a tight-binding model (for
the latter see \cite{FeIII}). The associated physical picture consists
of lightly or heavily doped semiconductors (e.g. Si doped with a
neighboring element which contributes excess electrons, for instance
P). This is the \textbf{Anderson model} (\cite{An1}, \cite{An2}),
described by the Hamiltonian 

\begin{eqnarray*}
H^{\omega }=\Delta + V^{\omega }  
\end{eqnarray*}
$$\eqno{(1)}$$
on 
$l^{2}(\mathbf{Z}^{d})$
where $\Delta$ is the (centered discrete Laplacian)
\begin{eqnarray*}
\left( \Delta u\right) _{n}=\sum_{n^{\prime }:\left\vert n-n^{\prime
}\right\vert =1}u_{n^{\prime }} 
\end{eqnarray*}
$$\eqno{(2)}$$
plus a perturbation by a random potential 
\begin{eqnarray*}
\left( V^{\omega }u\right) _{n}=V_{n}^{\omega }u_{n}
\end{eqnarray*}
$$\eqno{(3)}$$
where $\left\{ V_{n}^{\omega }\right\} _{n\in \mathbf{Z}^{d}}$ is a family
of independent, identically distributed random variables (i.i.d.r.v.) on a
probability space $\left( \Omega ,\mathcal{B},\mu \right) $, with a common
distribution $F(x)=\mu \left( \left\{ \omega :V_{n}^{\omega }\leq x\right\}
\right) $; $V_{n}^{\omega}$ is assumed to depend linearly on a quantity $v > 0$
(see the forthcoming (5b)), which is the disorder parameter, also
called coupling 
constant. The spectrum of $H^{\omega }$ is, by the ergodic theorem, almost
surely a nonrandom set $\sigma(H^{\omega })=[-2d,2d]+ \mbox{ supp } dF$
. Anderson conjectured that there exists a critical coupling
constant $0< v _{c}<\infty $ such that for $v \geq v _{c}$
the spectral measure of $H$ is pure point (p.p) for $\mu $--almost
every $\omega $, while, for $ v < v _{c}$ the spectral measure of $%
H^{\omega }$ contains two components, separated by so called
\textbf{mobility edge} $E^{\pm }$: if $E\in \lbrack E^{-},E^{+}]$
the spectrum of $H^{\omega }$ is pure absolutely continuous (a.c); in the
complementary set $\sigma (H^{\omega })\backslash \lbrack E^{-},E^{+}]$, $%
H^{\omega }$ has pure point spectrum - leading to the new important phenomenon of
\textbf{localization}, first proved for $d=1$ in \cite{GMP}, and for
$d \ge 3$ for the first time in by \cite{FMSS}, based on previous
seminal work by \cite{FS}, for fixed energy and large disorder, or for
fixed disorder and large energy (the latter meaning near the edges of
the band, i.e, the boundary of $\sigma(H^{\omega})$); see also
\cite{AM} for a much simpler proof by an entirely different method,
and references given there. 

As is well known, \textbf{resonance} is a phenomenon of cooperation
between two or more elements, while \textbf{randomness} is based on
independence - noncooperation - between elements. As remarked by
Howland \cite{Ho1}, it should not prove surprising, therefore, that
resonance in a system may be removed by randomizing parameters in it,
preventing, in the Anderson model, the cooperation necessary for
travelling waves (tunneling). What is, however, surprising, is the
(conjectured, see \cite{An2}) \textbf{sharpness} of the transition,
which means that for $v = v_{c}$ any slight (positive) variation of
the disorder parameter destroys the tunneling.   

This \textbf{tunneling instability} was studied by Jona-Lasinio,
Martinelli and Scoppola \cite{GMS} and Simon \cite{S4}, who showed
that tunneling is very sensitive to minimal changes in a double-well
potential, even those localized very far away from the minima ( a
phenomenon called ''flea on the elephant'' by Simon). Tunneling
instability may be of dynamical nature (see, e.g., \cite{GSa},
\cite{WC}), being also at the root of the existence of the chiral
superselection rules induced by the environment which account for the
shape of molecules such as Ammonia $NH_{3}$, see the review by Arthur
S. Wightman, "Superselection sectors: old and new",
\cite{Wightman}. We shall see that for a lattice model such as (1),
the idea that only a very ''slight'' perturbation of the Hamiltonian
has a drastic effect on the tunneling acquires even a new dimension in
the sparse situation described below. 

The proof of the existence of a mobility edge in the Anderson model
remains as one of the major open problems in mathematical
physics. Given the difficulties in proving this for the model (1)-(3),
one might be led to study the limit  
$ v \rightarrow 0$ of $H$, for which the spectrum is pure a.c.. We
shall instead follow a different approach to the Anderson conjecture
suggested by Molchanov (\cite{Mo} \cite{Mo1}, \cite{MoV}): the limit
of zero concentration, i.e., taking  $V^{\omega }$ in $H$ such that
\begin{eqnarray*}
V_{n}^{\omega }=\sum_{i}\varphi _{i}^{\omega }(n-a_{i})~,  
\end{eqnarray*}
with elementary potential ('' bump") $\varphi ^{\omega }:
\mathbb{Z}^{d}\longrightarrow \mathbb{R}$ satisfying a uniform integrability
condition
\begin{eqnarray*}
\left\vert \varphi ^{\omega }(z)\right\vert \leq \frac{C_{0}}{1+\left\vert
z\right\vert ^{d+\varepsilon }} 
\end{eqnarray*}
for some $\varepsilon >0$ and $0<C_{0}<\infty $ and
\begin{eqnarray*}
\lim_{R\rightarrow \infty }\frac{\#\left\{ i:\left\vert a_{i}\right\vert
\leq R\right\} }{R^{d}}=0\ .  
\end{eqnarray*}
$$\eqno{(4)}$$
Due to condition of zero concentration, potentials such as $V$ 
are called sparse and have been intensively studied in recent years since
the seminal work by Pearson \cite{Pe1} in dimension $d=1$ and by
Molchanov \cite{Mo}, \cite{Mo1} in the 
multidimensional case.  For $d\geq2$ the interaction between bumps is weak 
while for $d=1$ the phase
of the wave after propagation between distant bumps becomes 
''stochastic". This is the right moment to introduce our
one--dimensional model.

We consider (infinite) Jacobi matrices 
\begin{eqnarray*}
(J_{0}u)_{n}=u_{n+1}+ u_{n-1}\;,
\end{eqnarray*}
with the perturbation potential
$$(V^{\omega}u)_{n} = v^{\omega}_{n} u_{n} \eqno{(5a)}$$
with $u=(u_{n})_{n\geq 0}\in l_{2}(\mathbb{Z}_{+})$ ,and
\begin{eqnarray*}
v_{n}^{\omega}=\left\{ 
\begin{array}{lll}
v & \mathrm{ if } & n=a_{j}^{\omega }\in \mathcal{A}\,, \\ 
0 & \mathrm{ if } & \mbox{ otherwise }\,,
\end{array}
\right.  
\end{eqnarray*}
$$\eqno{(5b)}$$
for $v\in (0,1)$. 
$\mathcal{A}=\{a_{j}^{\omega }\}_{j\geq 1}$ denotes a random set of natural numbers
$a_{j}^{\omega }=a_{j}+\omega _{j}$ ,where $a_{j}$ satisfy the 
''sparseness condition'':
\begin{eqnarray*}
a_{j}-a_{j-1}=\beta ^{j}\;,\qquad \qquad j=2,3,\ldots 
\end{eqnarray*}
with $a_{1}+1=\beta \geq 2$ and $\omega _{j}$, $j\geq 1$,
are independent r.v. on a probability space
 $(\Omega ,\mathcal{B},\mu )$ uniformly distributed on a set
 $\Lambda_{j}=\{-j,\ldots ,j\}$. These variables introduce an
 uncertainty in the positions of 
those points for which $v_{n} \ne 0$: such models are called
\textbf{Poisson models} (see \cite{J} 
and references given there). Note that the support of the $\omega
_{j}$ only grows linearly with the suffix $j$. 
We write $J^{\omega} = J_{0} + V^{\omega}$, and denote by
 $J_{\phi }^{\omega }$ the operator associated to $J^{\omega }$ 
on the Hilbert space
 $\mathcal{H}$ of the square integrable sequences $u=\left(
u_{n}\right) _{n\geq -1}$ which satisfy a $\phi $-b.c. at $-1$:

\begin{eqnarray*}
u_{-1}\cos \phi -u_{0}\sin \phi =0  
\end{eqnarray*}
 
The essential spectrum of $J_{\phi}^{\omega}$ equals $[-2,2]$: it will
be represented as  
$\lambda = 2\cos \alpha$ with $\alpha \in [0,\pi)$. Zlatos \cite{Zl}
proved that this model exhibits a sharp 
transition from s.c. to p.p. spectrum. This was shown independently in
(\cite{CMW1},\cite{MWB}):  

\textbf{Theorem 1} Let $J_{\phi }^{\omega}$ be as above. Let
\begin{eqnarray*}
I\equiv \left\{ \lambda \in \lbrack -2,2]\backslash 2\cos \pi \mathbb{Q}:
\frac{1}{v^{2}}(\beta -1)(4-\lambda ^{2})>1\right\}
\end{eqnarray*}
with $v\in (0,1)$ and $\beta \in \mathbb{N}$, $\beta \geq 2$ . Then,
for almost all 
$\omega $ with respect to the uniform product measure on 
$\Lambda =\times_{j=1}^{\infty }\left\{ -j,\ldots ,j\right\} $,

(a) there exists a set $A_{1}$ of Lebesgue measure zero such that the
spectrum restricted to the set $I\backslash A_{1}$ is purely singular
continuous,

(b) the spectrum of $J_{P,\phi }$ is dense pure point when restricted
to  \hfill \break $I^{c}=\left( [-2,2]\backslash 2\cos \pi \mathbb{Q}\right)
\backslash I$ for almost every $\phi \in (0,\pi )$, where $\phi $
characterizes the boundary condition. Thus it is \textbf{purely}
p.p. in this interval.

\begin{figure}[ht!]
 \centering
\includegraphics[scale=0.7]{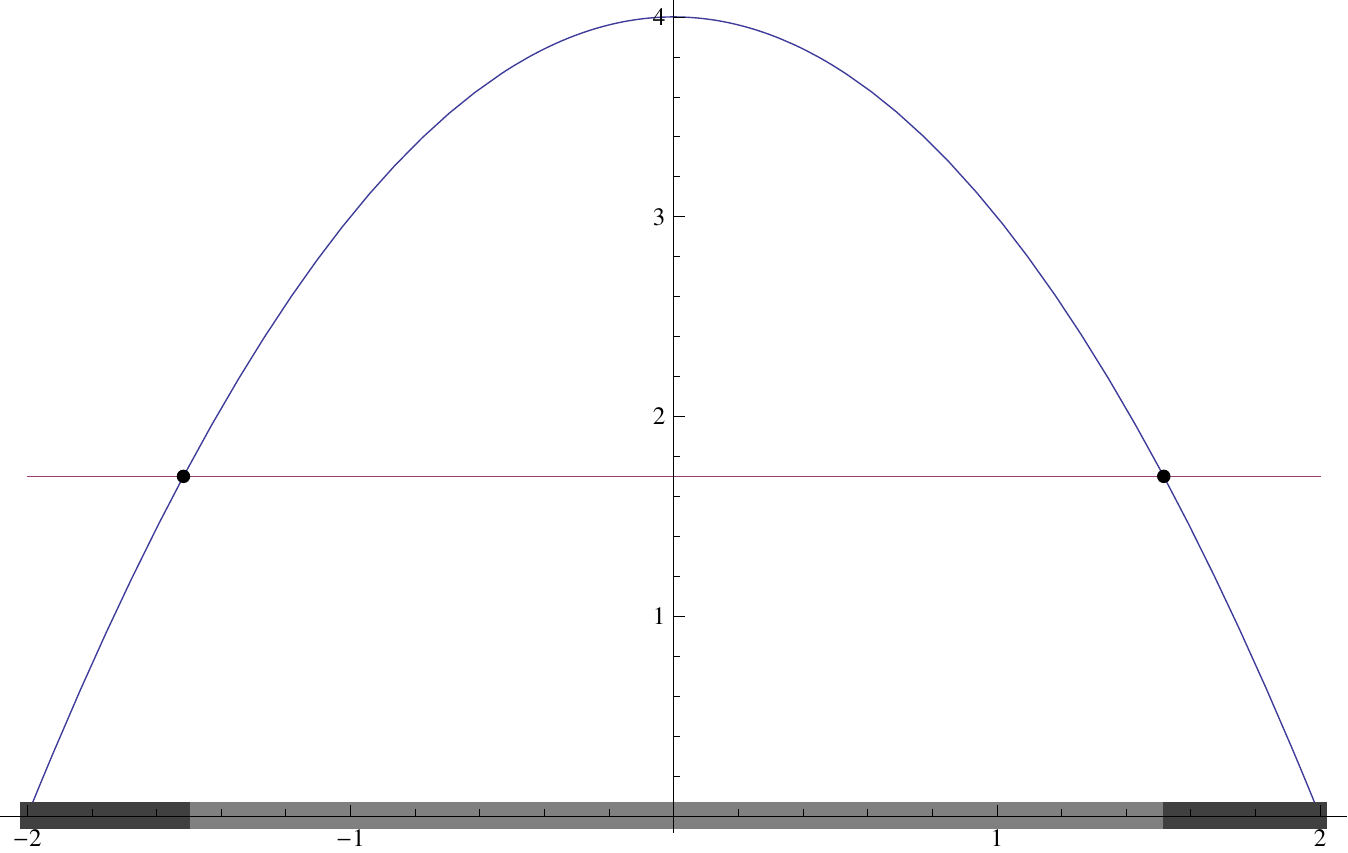}
\end{figure}

In the above figure, we see the s.c. (light gray) and p.p. (dark grey)
spectra separated by the ''mobility edges''  
$\lambda^{\pm} = \pm 2 \sqrt(1-\frac{v^{2}}{v_{c}})$, with
$\frac{v}{v_{c}}=1,3038 \cdots$. 

\textbf{Remark 1} 

The basic emphasis on one-dimensional (sparse) models has an important
technical reason: the profound approach of Gilbert and Pearson
connecting the space asymptotics of eigenfunctions of the restrictions
of a large class of Sturm-Liouville operators to finite intervals to
the spectral theory of these same operators in infinite space through
the concept of \textbf{subordinacy} \cite{GP} is only available in one
dimension. Using the important transfer matrix version of this theory
due to Last and Simon \cite{LaSi}, the surprise is that, even in a
regime of strong sparsity, a spectral transition from s.c. to
p.p. spectrum (first shown by Zlatos \cite{Zl}) may be proved: this is
theorem 1. Our proof (theorem 4.7 of \cite{MWB}) differs from that in
\cite{Zl} by the use of the (optimal) metrical version of Weyl's
theorem on uniform distribution \cite{KN} (in this case, of the
so-called Pr\"{u}fer angles, see \cite{KLS}) due to Davenport,
Erd\"{o}s and LeVeque \cite{DEL}. The latter permits an
\textbf{explicit} characterization of the exceptional set in which the
spectrum may not be of p.p. type (this exceptional set arises in
connection with the concept of essential support of a measure
\cite{GP}), which reveals it to be also of p.p. type. As a result, the
spectrum at the edges of the band is proved to be \textbf{purely}
p.p.. 

The robustness of the transition depicted in theorem 1 follows from
\cite{dR}, because the Hausdorff dimension of the s.c. spectrum may be
seen to be nonzero (\cite{Zl}, \cite{CMW3}).  

Finally, the difference Laplacean (2) leads to tunneling and bands
\cite{FeIII}, with a pure a.c. specrrum. The perturbation (5) which
leads to the emergence of p.p. spectrum is very slight on two
different counts: 1) the transition arises at a sharp (critical) value
$v_{c}$  of the disorder parameter $v$, 2) at the same time, we are in
a regime of strong sparsity! In this sense, the analogy to the ''flea
on the elephant'' \cite{S4} in a double-well potential is here much
sharper than in the originally conjectured  Anderson transition, and
the transition depicted in theorem 1, as well as in the higher
dimensional version in theorem 2, should indeed be viewed as \textbf{a
  concrete manifestation of tunneling instability}.

Unfortunately the s.c. spectrum does not posess either the dynamic or
the perturbation theoretic properties (see \cite{SWo}, \cite{Ho2}for
the latter) which are commonly associated with the physical picture of
delocalized states. Regarding the dynamical properties, for instance,
the \textbf{sojourn time} of a particle with initial state $\Psi$ in a
compact region  $S$ is defined as   
$$
J(S;\Psi) \equiv \int_{-\infty}^{\infty} ||P_{S} \exp(-itH) \Psi ||^{2} dt
$$
where $P_{S}$ denotes the projector on $S$. If $\Psi \in {\cal
  H}_{sc}$, it follows from a theorem by K.B. Sinha \cite{Si} that a
$S$ exists with $|S| < \infty$ and $J(S;\Psi) = \infty$. See also
\cite{MW}, \cite{MWB} for further discussion. 

We try therefore to attain higher dimensions.

\textbf{Multidimensional version}

Consider the Kronecker sum
$J_{\phi }^{\omega }$ as an operator on $\mathcal{H} \otimes \mathcal{H}$: 
\begin{eqnarray*}
J_{\theta }^{(2)}:=J_{\phi }^{\omega ^{1}}\otimes I+\theta I\otimes J_{\phi
}^{\omega ^{2}}  
\end{eqnarray*}
$$\eqno{(6)}$$ where $\omega ^{1}=\left( \omega _{j}^{1}\right)
_{j\geq 1}$ and $\omega 
^{2}=\left( \omega _{j}^{2}\right) _{j\geq 1}$ are two independent sequences
of independent random variables defined in $\left( \Omega ,\mathcal{B},\nu
\right) $, as before (we omit $\omega ^{1}$ and $\omega ^{2}$ in the l.h.s
above for brevity). Above, the parameter $\theta \in \lbrack 0,1]$
is included to avoid resonances . We ask for
properties of $J_{\theta }^{(2)}$ (e.g. the spectral type) which hold for 
\textbf{typical} configurations, i.e., a.e. $\left( \omega ^{1},\omega
^{2},\theta \right) $ with respect to $\nu \times \nu \times l$ where $l$ is
the Lebesgue measure in $[0,1]$. $J_{\theta }^{(2)}$ is a special
two--dimensional analog of $J_{\phi }^{\omega }$; if the latter was replaced
by $-\Delta +V$ on $L^{2}(\mathbb{R},dx)$ where $\Delta =d^{2}/dx^{2}$ is
the second derivative operator, and $V$ a multiplicative operator $V\psi
(x)=V(x)\psi (x)$ (potential), the sum above would correspond to $
\left( -d^{2}/dx_{1}^{2}+V_{1}\right) +\left( -d^{2}/dx_{2}^{2}+V_{2}\right) 
$ on $L^{2}(\mathbb{R}^{2},dx_{1}dx_{2})$, i.e., the
''separable case'' in two dimensions. Accordingly, we shall also refer to $
J_{\theta }^{(n)}$, $n=2,3,\ldots $, as the separable case in $n$ dimensions.

Our approach is to look at the quantity 

\begin{eqnarray*}
\left( \Phi ,e^{-itJ_{\theta }^{(2)}}\Psi \right) =f^{1}(t)f^{2}(\theta t)\\
\mbox{ where } \\
f^{i}(s)=f_{\mathrm{sc}}^{i}(s)+f_{\mathrm{pp}}^{i}(s)~,\quad i=1,2\\
\mbox{ with } 
f_{\mathrm{sc}}^{i}(s) &=&\int e^{-is\lambda }d\mu _{\varphi _{i},\psi
_{i}}^{\mathrm{sc}}(\lambda ) \\
f_{\mathrm{pp}}^{i}(s) &=&\int e^{-is\lambda }d\mu _{\rho _{i},\chi _{i}}^{
\mathrm{pp}}(\lambda )
\end{eqnarray*}

Above $\Phi ,\Psi \in \mathcal{H}\otimes \mathcal{H}$, 

\begin{eqnarray*}
\Phi &=&\left( \varphi _{1}\dot{+}\rho _{1}\right) \otimes \left( \varphi
_{2}\dot{+}\rho _{2}\right) ~,  \\
\Psi &=&\left( \psi _{1}\dot{+}\chi _{1}\right) \otimes \left( \psi _{2}\dot{
+}\chi _{2}\right) ~,
\end{eqnarray*}
with $\varphi _{i},\psi _{i}\in \mathcal{H}_{\mathrm{sc}}$, $\rho _{i},\chi
_{i}\in \mathcal{H}_{\mathrm{pp}}$ and $\varphi \dot{+}\rho $ denotes the
direct sum of two vectors $\varphi ,\rho \in \mathcal{H}$.

We shall use the following folklore proposition (see, e.g., theorem
3.2 of \cite{MWB} or \cite{Si}): 

\textbf{Proposition 1}

Let $\mu $ be a measure on the space $M(\mathbb{R})$ of all finite
regular Borel measures on $\mathbb{R}$. If the Fourier--Stieltjes transform
of $\mu $
\begin{eqnarray*}
\mathbb{R}\ni t\longmapsto \hat{\mu}(t)=\int e^{-it\lambda }d\mu (\lambda )
\end{eqnarray*}
belongs to $L^{2}(\mathbb{R},dt)$, then $\mu $ is absolutely continuous with
respect to Lebesgue measure.

The time-like decay of the Fourier-Stieltjes (F.S.) transform of the
spectral measure is dictated by the Hausdorff dimension of the
spectral measure: 

Let $S$ be a subset of $\mathbf{R}$, $\alpha \in [0,1]$, and $\delta > 0$. Define
\begin{eqnarray*}
h^{\alpha}_{\delta}(S) \equiv \inf \lbrace \sum_{j=1}^{\infty} |C_{j}|^{\alpha} \\
| S \subset \cup_{j=1}^{\infty} C_{j} \mbox{ with } |C_{j}| \le \delta \rbrace
\end{eqnarray*}
where $|C|$ denotes the Lebesgue measure (length) of $C$, and
$$
h^{\alpha}(S) = \lim_{\delta \to 0} h^{\alpha}_{\delta}(S) =
\sup_{\delta > 0} h^{\alpha}_{\delta}(S) 
$$
We call $h^{\alpha}$ $\alpha$ - dimensional Hausdorff measure on
$\mathbf{R}$. $h^{1}$ agrees with Lebesgue measure, and  
$h^{0}$ is the counting measure, so that $\{h^{\alpha} | 0 \le \alpha
\le 1\}$ is a family which interpolates continuously between the
counting measure and Lebesgue measure. 

\textbf{Definition 1} A Borel meaure on $\mathbf{R}$ is
\textbf{uniformly alpha-Hoelder continuous} (U$\alpha$H) if there
exists $C$ such that for every interval $I$ with $|I| < 1$, 
$$
\mu(I) < C |I|^{\alpha}
$$

By an important theorem of Last \cite{La2}, U$\alpha$H measures may be
obtained by a process of closure. In our one-dimensional model the
Hausdorff dimension varies locally in the s.c. spectrum, and the local
Hausdorff dimension (suitably defined) may be determined
explicitly. There exists a dense set in the s.c. subspace such that
the spectral corresponding spectral measure is U$\alpha$H. We shall
for simplicity assume that the Hausdorff dimension has a constant
value which will be denoted by $\alpha$. 

The basic theorem, due to Strichartz \cite{Str} (with a very slick
alternative proof by Last \cite{La2}, therefore we call it
Strichartz-Last theorem), relates decay in the \textbf{Cesaro sense}
to the Hausdorff dimension: 

\textbf{Strichartz-Last theorem} 

Let $\mu$ be a finite U$\alpha$H measure and, for each 
$f \in L^{2}(\mathbf{R}, d\mu)$, denote
$$
\hat{f\mu}(t) \equiv \int \exp(-ixt) f(x) d\mu(x)
$$
Then there exists $C$ depending only on $\mu$ such that for all $f \in
L^{2}(\mathbf{R},d\mu)$ and $T > 0$, 
$$
\langle |\hat{f\mu}|^{2} \rangle_{T} < C ||f||^{2} T^{-\alpha}
$$
where $||f||$ denotes the $L^{2}$ - norm of $f$, and $\langle g
\rangle_{T} \equiv \frac{\int_{0}^{T}g(t)dt}{T}$. 

A measure is called a \textbf{Rajchman measure} iff
$$
\lim_{|t| \to \infty} \hat{\mu}(t) = 0
$$
It does \textbf{not} follow from the decay in the Strichartz-Last
theorem that the corresponding measure is Rajchman . Let $E$ denote
the usual Cantor ''middle-thirds'' set in $[-\pi,\pi]$; die
F.S. transform of the corresponding measure $\Gamma$ is: 
$$
\Gamma(u) = \prod_{j=1}^{\infty} \cos[2/3 u \pi 3^{-j+1}]
$$
The corresponding Hausdorff dimension is well-known to be
$\alpha= \frac{|\log 1/2|}{|\log 1/3|}$ und the measure is U$\alpha$H,
but, from the above formula for $\Gamma$ it follows that 
$$
\Gamma(n) = \Gamma(3n) \mbox{ for all } n \in \mathbf{Z}
$$
and hence the Cantor measure is not Rajchman.

We are now ready to state our main result:

\textbf{Theorem 2}
Let
\begin{eqnarray*}
v^{2}<a\left( \sqrt{\beta }-1\right) <v_{c}^{2}
\end{eqnarray*}
with $a<4$, where $v_{c} = 2\sqrt(\beta -1)$. Then, for almost every
$\left( \omega ^{1},\omega ^{2},\theta\right) $ with respect to $\nu
\times \nu \times l$, 
[a.] there exist $\tilde{\lambda}^{\pm }$ with $\tilde{\lambda}^{+}=-
\tilde{\lambda}^{-}$ and 
\begin{eqnarray*}
0<\tilde{\lambda}^{+}<\lambda ^{+} 
\end{eqnarray*}
such that
\begin{eqnarray*}
\left( \tilde{\lambda}^{-}(1+\theta ),\tilde{\lambda}^{+}(1+\theta )\right)
\subset \sigma _{\mathrm{ac}}\left( J_{\theta }^{(2)}\right)  
\end{eqnarray*}

[b.]
\begin{eqnarray*}
\left[-2(1+\theta ),\lambda ^{-}(1+\theta )\right) \cup \left( \lambda
^{+}(1+\theta ),2(1+\theta )\right] \subset \sigma _{\mathrm{pp}}\left(
J_{\theta }^{(2)}\right)
\end{eqnarray*}

[c.]
\begin{eqnarray*}
\sigma _{\mathrm{sc}}\left( J_{\theta }^{(2)}\right) \cap \left( \lambda
^{-}(1+\theta ),\lambda ^{+}(1+\theta )\right)  
\end{eqnarray*}
may, or may not, be an empty set.

The basic idea of the proof is that the Strichartz-Last theorem
suggests a pointwise decay of the F.S. transform of the spectral
measure of type $t^{-\alpha/2}$ for large $t$. For the Kronecker sum
(with $\theta=1$) the F.S. transform is the product of the
corresponding transforms for the one-dimensional system, which we
expect decays as $t^{-\alpha}$. In order to use the proposition, we
restrict the spectral measure to an interval in the s.c. spectrum such
that $2\alpha > 1$ which is in principle non-empty (and may be proven
so), and denote the F.S. transform of the thus restricted measure by
the same symbol as before. It turns, however, out that this heuristics
is not correct mathematically: as shown above, Cesaro decay
\textbf{does not} in general imply pointwise decay or, in other words,
that the spectral measure is Rajchman. This is a much harder problem -
(see \cite{MWB}, chapter 4 for the full treatment of the pointwise
decay of a model with superexponential sparsity and pure
s.c. spectrum, and the recent very hard analysis to prove local decay
in nonrelativistic QED \cite{CJFS}). Moreover, the result for general
$\theta$ may not be true due to resonance between Cantor sets, a
subtle phenomenon which has been so far analysed only in the
self-similar case by Hu and S.G. Taylor \cite{HuT}. Self-similarity
occurs, however, seldom: indeed, the s.c. spectrum of sparse Jacobi
matrices is not self-similar by a theorem of Combes and Mantica
\cite{CM}. We proved, however, that the main idea is indeed correct by
generalizing a method due to Kahane and Salem (\cite{KaS1},
\cite{KaS2}) in specific (self-similar) cases to the problem at hand. 

\textbf{Conceptual relevance of the present model: comparison with the
Anderson model on a Cayley tree}

Sparse models in dimension $d\ge2$ may be good models for the Anderson
transition in lightly doped semiconductors 
( which takes place already for $d=2$ ): for the latter see
\cite{ShEf}. It seems natural to expect that 
the present model might pave the way for a good qualitative description of
the Anderson transition in lightly doped semiconductors. We say
'' pave the way" because the present form of the model is not
adequate for a physical description -- but we argue
that the main objection may be eliminated by considering a truly $d$
--dimensional model.

The main objection is, of course, that exponential sparsity  is
too severe, and not physically reasonable. It must be recalled, however,
that the separable model does not take account of dimensionality in a proper
way, because ''truly" three dimensional sparse models may drastically
change, the cardinality of $\left\{ i:\left\vert
a_{i}\right\vert \leq R\right\} $ in (4) from $O\left( \log R\right) $ to 
$O(R^{d-\varepsilon })$ in dimension $d$, for some $\varepsilon >0$, which is
still compatible with the assumption of sparsity, changing, at the
same time, the conditions on the sparsity for the existence of the transition.

It should be useful to compare the present model with the Anderson
model on a Cayley tree (\cite{Kl}, \cite{J}, \cite{AiWa1},
\cite{AiWa2}). A Cayley tree (or Bethe lattice) is a graph with no
closed loops: there is a central site and three generations in a
regular tree of ramification $r=2$. All sites are always connected to
$r+1$ nearest neighbors. For a tree with $N$ generations, there are
$N_{s} = (r+1)r^{N-1}$ sites at the surface (generation $N=0$) and a
total  $N_{T} = (r^{N+1} + r^{N-2})/(r-1)$ sites. Since the ratio
$N_{S}/N_{T}$ is nonzero for $N \to \infty$, the behaviour of
statistical systems on a Cayley tree is usually (but see later)
pathological and quite distinct from the physical features of a
Bravais lattice. One notable example is the \textbf{a.c. spectrum}
found in the Anderson model on a Bethe lattice in the seminal work of
A. Klein \cite{Kl} (see also \cite{J} for a review and further
references): the extended states decay exponentially but are not
square-integrable due to the $\exp(n)$ many points within $n$ links of
a given vertex alluded to before, and is thus of \textbf{entirely
  different nature} from the a.c. spectrum found on a Bravais lattice
- and in particular in theorem 2. 

It is often mentioned in the literature that the Bethe lattice
describes the \textbf{infinite-dimensional} limit of a hypercubic
lattice. Indeed, the Bethe or Bethe-Peierls approximation (BPA)
(\cite{KHu}, pg. 357), which improves on the mean field approximation
(which describes the infinite-dimensional limit) by taking into
account specific short-range order, agrees with the Bethe lattice in
the following sense \cite{CJT}: expectation values of spins ''far
removed from the surface'' on a Cayley tree agree with those obtained
by the BPA. That this is truly only a \textbf{local} property is made
clear by the fact that, by first considering a finite Bethe lattice
and then performing the thermodynamic limit one does \textbf{not}
obtain the BPA (see again the remarks in \cite{CJT}). 

Klein's proof of the Anderson transition \cite{Kl} uses the loopless
character of the graph in an essential way and thus, as remarked by
Jitomirskaya \cite{J}, the Bethe lattice, while infinite-dimensional
(but in the above-mentioned sense), is, in a sense,
one-dimensional. Similarly, our model has no loops, because theorem 2
"inherits'' the one-dimensional structure of the model in theorem
1. The geometrical structure underlying (6), i.e., the three
coordinate half-axes (there would be no problem in extending it to
cover the three full axes) is, however, a (small!) part of the full
(sparse) model on a Bravais lattice. Thus, in our opinion, in spite of
the considerable independent mathematical interest of the recent work
on the Anderson model on a Bethe lattice (\cite{AiWa1}, \cite{AiWa2}),
the present model is physically more reasonable (for the description
of the Anderson transition in lightly doped semiconductors). Of
course, the flaws discussed at the beginning of this subsection will
only be disposed of by ''filling in'' the ''remaining'' points -an
immensely challenging, basic open problem!. 

As in the Bethe lattice \cite{Kl}, the sharpness of the transition,
i.e., the existence of a mobility edge, was not proved for the present
model. Recent work \cite{AiWa1} proves that no mobility edge occurs in
the Bethe lattice at weak disorder. The general character of the
arguments used in theorem 2 to establish the existence of
a.c. spectrum (see the sketch of the ideas of the proof there and
\cite{MW}) suggests that the intermediate region might be more
accessible to analysis than the Bethe lattice, but this remains a
(challenging) open problem. 

It is, however, rewarding that already the separable model displays a
dramatic ''kinematic'' effect of the dimensionality: for $ d \ge 2$
the transition becomes truly Anderson-like, i.e., from a.c. to
p.p. spectrum. we say dramatic because, in the case of heavy doping,
the $d \ge 2$ version, built as in (6) from the one-dimensional
version of (1) - (3), \textbf{continues} to have purely p.p. spectrum,
by \cite{GMP} and the same proof of theorem 2 (more precisely, the
easy part of theorem 2), in complete disagreement with the expected
transition! Thus, theorem 2 is a definite indication that the present
approach via ''light doping'' is more likely to produce a transition
in the full (sparse) version.  

\textbf{Frustration and short-range spin glasses: the Edwards-Anderson model}

Dilute solutions of atoms of large magnetic moments (such as the
transition metals Fe, Co, Mn) in a paramagnetic substrate (Cu, Au)
present a number of peculiar physical properties. For small, but
sufficiently high concentrations of magnetic impurities, the
susceptibility in low fields displays a characteristic peak, with
discontinuous derivatives, at a temperature $T_{g}$. The specific heat
is always smooth, with a linear dependence on the temperature as $T
\to 0$. The behaviour of these \textbf{spin glasses} has been
explained in terms of an indirect RKKY interaction between the spins
of these magnetic impurities mediated by the electrons of the
paramegnetic matrix \cite{BiYo}, which is of the form
$-J(|i-j|)S_{i}S_{j}$ with  
$J(r) = (k_{F}r)^{-3}\cos(2k_{F}r)$, where $r$ is the distance between
magnetic atoms and $k_{F}$ the fermi momentum;  
$S_{i}=\pm 1$ are (e.g.) Ising spins. The rapid oscillations and the
weak decay of $J(r)$, as well as the random distribution of the
impurity magnetic atoms, are the basic ingredients of spin glasses. At
sufficiently low temperatures there is a ''freezing'' of the magnetic
moments in random directions (which leads to an increase of the
susceptibility). The spin glass phase may be regarded as a
conglomerate of blocks of spins, each block with its own
characteristic orientation, in such a way that there is no macroscopic
magnetic moment \cite{BiYo}. 

Edwards and Anderson \cite{EA} proposed a spin Hamiltonian, which we
write in a generalized version as follows. 
For each finite set of points $\Lambda \subset \mathbf{Z}^{d}$, where
the dimension $d$ will be restricted to the values $d=2$ and $d=3$,
consider the Hamiltonian 
$$
H_{\Lambda}(\{J\}) = \sum_{i,j \in \Lambda} J_{i,j} \Phi_{i,j}
\eqno{(7a)}
$$
where $\Phi_{i,j}$ with $i,j \in \Lambda$ are self-adjoint elements of
the algebra generated by the set of spin operators, the Pauli matrices
$\sigma_{i}^{x}$, $\sigma_{i}^{y}$, $\sigma_{i}^{z}$, $i \in \Lambda$,
on the Hilbert space ${\cal H}_{\Lambda} = \otimes_{i \in \Lambda}
\mathbf{C}_{i}^{2}$, given by 
$$
\Phi_{i,j} = \alpha_{x} \sigma_{i}^{x}\sigma_{j}^{x} + \alpha_{y}
\sigma_{i}^{y}\sigma_{j}^{y}  
+\alpha_{z}\sigma_{i}^{z}\sigma_{j}^{z}
\eqno{(7-b)}
$$
for $|i - j| = 1$, and zero otherwise. The random couplings $J_{i,j}$,
with $|i - j| = 1$, are random variables (r.v.) on a probability space
$(\Omega, {\cal F}, P)$, where ${\cal F}$ is a sigma algebra on
$\Omega$ and $P$ is a probability measure on $(\Omega,{\cal F})$. We
may take without loss of generality  
$$\Omega = \times_{B^{d}} S \eqno{(8)}$$ where $S$ is a Borel subset
of $\mathbf{R}$, $B^{d}$ is the set of bonds in $d$ dimensions, and
assume that the $J_{i,j}$ are independent, identically distributed
r.v.. In this case, $P$ is the product measure $$dP = \times_{B^{d}}
dP_{0} \eqno{(9)}$$ of the common distribution $P_{0}$ of the random
variables, which will be denoted collectively by $J$. The
corresponding expectation (integral with respect to $P$) will be
denoted by the symbol $Av$. We have to assume that $$ Av(J_{i,j}) = 0
\eqno{(10)}$$ for all $i,j \in \mathbf{Z}^{d}$, i.e., that the
couplings are centered. This assumption mimicks the rapid oscillations
of the RKKY interaction. Let $E_{\Lambda}$ denote the GS energy of
$H_{\Lambda}$, i.e., $E_{\Lambda} = \inf spec(H_{\Lambda})$. The
following result was proved, among several others, in \cite{ConL}: 

\textbf{Theorem 3} (\cite{ConL}) For $P$- a.e. $\{J\}$, the limit
below exists and is independent of the b.c.: 
$$
e^{(d)} \equiv \lim_{\Lambda \nearrow \mathbf{Z}^{d}} \frac{E_{\Lambda}}{|\Lambda|}
\eqno{(11-a)}
$$
and
$$
e^{(d)} = \inf_{\Lambda} \frac{E_{\Lambda}}{|\Lambda|}
\eqno{(11-b)}
$$
where $|\Lambda|$ denotes the number of sites in $\Lambda$. Finally,
$$
e^{(d)} \ge e^{(d+1)}
\eqno{(11-c)}
$$

(11-a) is the far-reaching property of self-averaging (see \cite{An3}
for a discussion): it expresses that measurable - e.g. thermodynamic-
quantities are the same for any \textbf{typical} configuration of the
sample, i.e., are experimentally reproducible. It follows from (11a)
that, $P$- a.e., 
$$
e^{(d)} = \lim_{\Lambda \nearrow \mathbf{Z}^{d}} \frac{Av(E_{\Lambda})}{|\Lambda|}
\eqno{(11-d)}
$$
Let $\Lambda_{N}$ denote a square with $N$ sites if $d=2$ or a cube
with $N$ sites if $d = 3$, and write  
$H_{N}^{(d)}(J) \equiv H_{\Lambda_{N}}(J)$. We now adopt periodic
b.c. for simplicity, but the final result is independent of the
b.c. due to theorem 1. We may write 
$$
H_{N}^{(d)}(J) = \sum_{n \in \Lambda_{N}} H_{n}^{(d)}(J)
\eqno{(12-a)}
$$
where $H_{n}^{(d)}$ is given by
$$
H_{n}^{(d)}(J) = c_{d} \sum_{(i,j) \in \Lambda_{n}^{(d)}} J_{i,j}\Phi_{i,j} 
\eqno{(12-b)}
$$
Above, $c_{d}$ are factors which eliminate the multiple counting of bonds, i.e.,
$$
c_{2} = 1/2
\eqno{(12-c)}
$$
and
$$
c_{3} = 1/4
\eqno{(12-d)}
$$
and $\Lambda_{n}^{(2)}$ is a square labelled by a site $n$, for which
we adopt the convention, using a right-handed $(x,y)$ coordinate
system, that $n=(n_{x},n_{y})$ is the vertex in the square with the
smallest values of $n_{x}$ and $n_{y}$. Similarly, $\Lambda_{n}^{(3)}$
is a cube labelled by a site $n = (n_{x},n_{y},n_{z})$ with, by the
same convention, the smallest values of $n_{x}$, $n_{y}$ and
$n_{z}$. Due to the periodic b.c., the sum in (12-b) contains
precisely $N$ lattice sites. The notation $H_{n}^{(d)}$ is short-hand
for its tensor product with the identity at the complementary lattice
sites in  
$\Lambda_{N}\backslash \Lambda_{n}^{(d)}$.
Let $E_{0}^{(N,d)}(J)$ denote the GS energy of $H_{N}^{(d)}(J)$ ,
$E_{0}^{(n,d)}(J)$ the GS energy of $H_{n}^{(d)}(J)$ and
$$
E_{0}^{(d)} \equiv Av(E_{0}^{(n,d)}(J))
\eqno{(12-e)}
$$
By the condition of identical distribution of the r.v. $J_{i,j}$,
$E_{0}^{(d)}$ does not depend on $n$, which is implicit in the
notation used. We have 

\textbf{Theorem 4} (Theorem 2 and proposition 1 of \cite{Wr5}) The
following lower bound holds: 
$$
\frac{Av(E_{0}^{(N,d)}(J))}{N} \ge E_{0}^{(d)}
\eqno{(13-a)}
$$
Further, let $dP_{0}$ in (9) be the Bernoulli distribution $dP_{0} =
1/2(\delta_{J} + \delta_{-J})$, set for simplicity $J = 1$ and let
$\alpha_{x} = \alpha_{y} = 0$ in (7b). Then, 
$$
e^{(2)} \ge -3/2
\eqno{(13-b)}
$$
and
$$
e^{(3)} \ge -1/4\frac{36096}{4096}
\eqno{(13-c)}
$$

The special case $\alpha_{x} = \alpha_{y} = 0$ in (7-b) is the
classical EA spin-glass; we also set $\alpha_{z} = 1$. In this case
$e^{(d)}$, given by the r.h.s. of (11-d), is invariant under the
''local gauge transformation''  
$\sigma_{i}^{z} \to -\sigma_{i}^{z}$ together with $J_{i,j} \to
-J_{i,j} \mbox{ for all } j | |j-i|=1$, whatever the lattice site $i$.  

According to the above, \cite{To}, an elementary square
(''plaquette'') $P$ is said to be \textbf{frustrated}
(resp. non-frustrated) if $G_{P} \equiv \prod_{(i,j) \in P} J_{i,j} =
-1$ resp. $+1$. Note that $G_{P}$ is gauge-invariant and that, for the
quantum XY (or XZ) model defined by setting $\alpha_{y} = 0$ in (7-b),
$e^{(d)}$ is also locally gauge-invariant if we add to the above
definition the transformation $\sigma_{i}^{x} \to -\sigma_{i}^{x}$,
i.e., the transformation on the spins is defined to be a rotation of
$\pi$ around the $y$ - axis in spin space. The property of local
gauge-invariance guarantees the absence of a macroscopic magnetic
moment (spontaneous magnetization) mentioned before as a basic
property of spin glasses \cite{ARS}. 

Since the Pauli z-matrices commute, finding the minimum eigenvalue of
(7-a) in the classical EA case is equivalent to find the configuration
of Ising spins $\sigma_{i} = \pm 1$, denoted collectively by $\sigma$,
which minimizes the functional 
$$
F(\sigma,J) \equiv \sum_{\sigma} J_{i,j} \sigma_{i} \sigma_{j}
$$ 
The minimal energy of a frustrated plaquette equals $E_{P,f} = -2$ and
of a non-frustrated plaquette $E_{P,u} = -4$.  

(13b,c) of theorem 4 provide the first (nontrivial) rigorous lower
bound both for $e^{(2)}$ and $e^{(3)}$. Using the natural misfit
parameter $$m = \frac{|E^{id}| - |E_{0}|}{|E^{id}|} \eqno{(14)}$$ as a
measure of plaquettes frustrated or bonds unsatisfied (see (4) of
\cite{KK}), where $E_{0}$ denotes the ground state energy of the
frustrated system and $E^{id}$ is the ground state energy of a
relevant unfrustrated reference system, we find from (13b) in the $d =
2$ case the lower bound $m \ge 0.25$ and for $d = 3$, from (13c), the
lower bound $m \ge 0.26\cdots$: thus, in both cases, the measure of
frustrated plaquettes or unsatisfied bonds as defined above is at
least of the order of $1/4$. 

The method of proof of theorem 4 is a rigorous version of a
finite-size cluster method, originally due to \cite{BaS}, together
with the variational principle. If we take in (7) $\alpha_{z} = 1$,
$\alpha_{y} = 0$, and $\alpha_{x} = \alpha$, and consider $\alpha$ as
a small parameter, we have the anisotropic XZ (or XY) model. By the
norm-equicontinuity (in the volume) of $H_{\Lambda}(\{J\})$ (given by
(7)) as a function of $\alpha$, which is preserved upon taking
averages over the probability distribution of the $J$, it follows that
(13b,c) hold with the right hand sides varying by small ammounts if
$|\alpha|$ is sufficiently small: this is conceptually important for
reasons mentioned in the introduction. 

The mean field theory, recently rigorously solved in a spectacular way
(see \cite{Gue} for a review) does not exhibit frustration - indeed,
this concept is not even generally defined for their model, since the
theory does not require an underlying lattice. Whether frustration is
an important issue in the description of realistic spin glasses is an
important open problem. 

We refer to the article by Bovier and Fr\"{o}hlich \cite{BoFr} (see
also Bovier's book \cite{Bo}) for an illuminating discussion of
complementary, mostly global (i.e., involving the lattice as a whole)
aspects of frustration. Our bound (13c), which relies only on the
\textbf{local} structure, is, however, slightly better than
Kirkpatrick's \cite{SKir}, which is based on very reasonable, but
unproved conjectures of a global nature. A most relevant open problem
would be, of course, to extend the finite-size cluster method to
obtain bounds for the free energy of the EA model. 

\textbf{Return to equilibrium and (thermodynamic) instability of the first and second kinds}

The dynamics of spin glasses is a topic of great relevance, both
conceptual and experimental. It happens, however, that the standard
approach to the kinetic theory (e.g., for the mean-field spin glass
model) relies on Glauber dynamics (see, e.g., \cite{Sz}),a well-known
dynamics imposed on the Ising model, which has been used to study
metastability in early days \cite{CCO} and more recently
\cite{SchS}. In spite of the considerable independent mathematical
interest of the latter works, it should be remarked that only for a
quantum system does a physically satisfactory definition of the
dynamics of the states and observables exist which is relevant to the
microscopic domain, including, of course, condensed matter
physics. The scarcity of examples of approach to equilibrium in
quantum mechanics is due, of course, to the extreme difficulty of
estimating  specific properties of the quantum evolution.  

This fact was the basic motivation which prompted Emch \cite{Em} and
Radin \cite{Ra} to propose a quantum dynamical model (which we dubbed
the Emch-Radin model in \cite{Wr5} and \cite{MWB}), which is relevant
to systems with high anisotropy, and displays a remarkable property of
\textbf{non-Markovian approach to equilibrium}, or return to
equilibrium. It turns out that such a model is useful not only for a
description of nuclear spin-resonance experiments - such as the one
\cite{LNo} which motivated Emch and is described below - but also to
describe dynamical effects associated to quantum crossover phenomena
in spin glasses (see \cite{Sach} for a review and references). For
this purpose, a material was chosen with a strong spin-orbit coupling
between the spins (of the magnetic ion) and the underlying crystal:
this coupling essentially restricts the spins to orient either
parallel or antiparallel to a specific crystalline axis, which we
shall label as the z-axis. Such spins are usually referred to as Ising
spins, and will be described by the Ising part of the forthcoming
Hamiltonian. In the experiments (see \cite{Sach} and references), a
transverse magnetic field is then applied, oriented perpendicular to
the z axis. A large enough transverse field will eventually destroy
the spin glass order even at $T=0$, leding to the existence of a
crossover region. 

We shall be interested in a situation in which a large transverse
field has been applied; the \textbf{initial} state of the system is,
then, approximately, a product state of the forthcoming form (16). The
same experimental setup should therefore allow a measurement of the
rate of \textbf{return to equilibrium} of the mean transverse
magnetization, i.e., of how $f(t)$, defined by (20), tends to zero, or
of how (21) is approached. As we shall see, this rate depends strongly
on the probability distribution of the $J$.    

The XY model has also been extensively studied from the point of view
of return to equilibrium, since early days as an example of general
theory \cite{Ro}, more recently in \cite{AB} (see the latter for
reference to related important work of Araki), but unfortunately only
in one dimension.

\textbf{The nonrandom model}

Following Radin's review \cite{Ra} of the work of Emch \cite{Em},
consider the experiment of \cite{LNo}: a $CaF_2$ crystal is placed in
a magnetic field thus determinig the z-direction, and allowed to reach
thermal equilibrium. A rf pulse is then applied which turns the net
nuclear magnetization to the $x$ direction. The magnetization in the
$x$ direction is then measured as a function of time.  

As in (\cite{Em},\cite{Ra}), we assume an interaction of the form
$$
H_{V}=1/2\sum_{j,k}\epsilon(\vert j-k \vert)\sigma^{z}_{j}\sigma^{z}_{k} 
\eqno{(15)}
$$
where $j\in V$ and $k\in V$ in (15), with $V$ is a finite region,
$V\in\mathbf{Z}^{d}$. $H_{V}$ is defined on the Hilbert space ${\cal
  H}=\bigotimes_{i\in V} \mathbf{C}^{2}$. The state representing the
system after the application of the rf pulse will be assumed to be the
product state 
$$
\rho=\rho(0)=\bigotimes_{j\in V}\phi_{j}
\eqno{(16-a)}
$$

where

$$
\phi_{j}(\cdot)=tr_{j}(\cdot \exp(-\gamma\sigma^{x}_{j}))/tr
_{j}(\exp(-\gamma\sigma^{x}_{j}) 
\eqno{(16-b)}
$$
and $tr$ is the trace on $\mathbf{C}^{2}_{j}$. Other choices of the
state are possible \cite{Ra}, but we shall adopt (16) for
definiteness. Let 
$$
S^{x}=1/N(V)\sum_{j\in V}\sigma^{x}_{j}
\eqno{(17)}
$$
be the mean transverse magnetization, with $N(V)$ denoting the number
of sites in $V$. The real number $\gamma$ in may be chosen as in
\cite{Em} such as to maximize the microscopic entropy subject to the
constraint $tr S^{x}\rho(0)$ equal to a constant, i.e., a given value
of the mean transverse magnetization. Since the state (16) is a
product state, 
$\gamma$ is independent of $V$, and has the same value if $S^{x}$ is
replaced by $\sigma^{x}_{i_0}$, for any $i_0\in V$. Let 
$$
\rho^{V}_{t}\equiv U_{V}^{t} \rho(0)U_{V}^{-t}
\eqno{(18-a)}
$$
and
$$
\langle\sigma^{x}_{i_0}\rangle_{V}(t)\equiv \rho^{V}_{t}(\sigma^{x}_{i_0})
\eqno{(18-b)}
$$
where $U_{V}(t)=\exp(itH_{V})$. We may write, by (18),
$$
\langle\sigma^{x}_{i_0}\rangle_{V}(t)=\rho(0)(U_{V}(-t)\sigma^{x}_{i_0}U_{V}(t))
\eqno{(18-c)}
$$
It is natural to define
$$
\langle\sigma^{x}_{i_0}\rangle(t)=\lim_{V\to\infty}\langle\sigma^{x}_{i_0}\rangle_{V}
(t)     
\eqno{(19)}
$$
provided the limit on the r.h.s. of (19) exists, as the expectation
value of the local transverse spin in the (time-dependent)
nonequilibrium state $\rho_{\infty}^{t}$ of the infinite system. A
weak-star limit of the sequence of states  
$\rho_{V}^{t}(\cdot)= tr(\cdot U_{V}^{t}\rho(0)U_{V}^{-t})$ exists, by
compactness, on the usual quasilocal algebra  
$\cal{A}$  of observables (see, e.g. \cite{Se3}, for a comprehensive
introduction to the algebraic concepts employed here). The expectation
value of $\sigma^{x}_{i_0}$ in the equilibrium state associated to
(15) is zero by the symmetry of rotation by $\pi$ around the $z$
axis. We may now pose the question whether the limit 
$$
f(t)\equiv
\lim_{N(\Lambda)\to\infty}\rho_{\infty}^{t}(1/N(\Lambda)\sum_{i_0\in\Lambda}
\sigma^{x}_{i_0})   
\eqno{(20)}
$$
where $\Lambda$ denotes a finite subset of $\mathbf{Z}^{\nu}$, which
is interpreted as the mean transverse magnetization, exists. The
property of return to equilibrium is expressed  by  
$$
\lim_{t\to\infty}f(t)=0
\eqno{(21)}
$$  
Of particular interest is the rate of return to equilibrium. In the
present nonrandom case, the limit at the r.h.s. of (21) equals
$f(t)=\rho_{\infty}^{t}(\sigma^{x}_{i_0})$, for any $i_0$, by
translation invariance of $\rho_{\infty}^{t}$. This is not so for
random systems, in which case additional arguments are necessary to
show the convergence of the r.h.s. of (21) for almost all
configurations of couplings. 

We assume that the function $\epsilon(\cdot)$ satisfies:
$$
\epsilon(0)=0
\eqno{(22)}
$$
and either:
$$
\sum_{j \in \mathbf{Z}^{d}}  \epsilon(j)  < \infty \mbox{ i.e. }
\epsilon\in l_1(\mathbf{Z}^{d}) 
\eqno{(23-a)}
$$
or
$$
\sum_{j \in \mathbf{Z}^{d}} (\epsilon(j))^{2} < \infty  \mbox{ i.e. }
\epsilon\in l_2(\mathbf{Z}^{d}) 
\eqno{(23-b)}
$$

\textbf{ The random model}

We now introduce the Hamiltonian of a disordered system corresponding
to (15), where we take, for simplicity,  
$$V = V_{n} = [-n,n]^{d} \eqno{(23-d)}$$ :
$$
\tilde{H_{n}}=1/2\sum_{j,k \in V_{n}} J_{(j,k)}\epsilon(j-k)\sigma^{z}_{j}\sigma^{z}_{k}
\eqno{(24)}
$$
where $J_{(j,k)}$ are independent, identically distributed random
variables (i.i.d. r.v.) on a probability space which we denote by
$(\Omega,{\cal B}, P)$. We shall use  $Av(\cdot)$ to denote averaging
with respect to the random configuration $\{J_{j,k}\}$, which we
denote collectively by the symbol $J$. The $J$ are assumed to satisfy
\cite{vEvH}: 
$$
Av(J_{(j,k)})=0
\eqno{(25-a)}
$$
$$
| Av(J^{n}_{(j,k)})| \le n!c^{n} \forall n=2,3,4,\cdots
\eqno{(25-b)}
$$
We shall take, without loss, $\epsilon(j) \ge 0$. If
\begin{eqnarray*}
\epsilon(j) = \left\{
\begin{array}{ll}
\beta & \mbox{ if } j \in \pm \delta_{i}\,,\\
0     & \mbox{ otherwise }\,,
\end{array}
\right.
\end{eqnarray*}
$$\eqno{(25-c)}$$
where $\pm \delta_{i}$ for $i=1, \cdots, d$ denote the set of $z=2d$
bonds connecting the origin to a point of $\mathbf{Z}^{d}$, we have
the EA spin glass model of section 3.  

Let $|V_{n}|$ denote the number of sites in $V_{n}$ and the free
energy per site $f_{n}$ be defined by 
$$
f_{n}(J)\equiv \frac{-kT\log Z_{n}(J)}{|V_{n}|}
\eqno{(26-a)}
$$
where
$$
Z_{n}(J)= tr(\exp(-\beta\tilde{H_{n}}))
\eqno{(26-b)}
$$
is the partition function and the trace is over the Hilbert space
${\cal H}$. Then 

\textbf{Theorem 5}\cite{vEvH} (see also \cite{KhSi} for the first
result in this direction):

Under assumptions (12-b) and (25), the thermodynamic limit of the free
energy per site 
$$
f(J)=\lim_{n\to\infty} f_{n}(J)
\eqno{(27-a)}
$$
exists and equals its average:
$$
f(J)= Av(f(J))=\lim_{n\to\infty} Av(f_{n}(J))
\eqno{(27-b)}
$$
for almost all configurations $J$ ($\mbox{ a.e. } J $).

The reason why (23-b) suffices for the existence of the thermodynamic
limit is that, in order to obtain a uniform lower bound for the
average free energy per site, the cumulant expansion (see, e.g.,
(\cite{S6}, (12.14), pg 129, for the definition of $Av_c$, there
called Ursell functions): 
\begin{eqnarray*}
Av(\exp(tJ_{(i,j)})= \exp(\sum_{n=2}^{\infty}Av_c(J^{n}_{(i,j)})t^{n}/n!)
\end{eqnarray*}
was used \cite{vEvH}, which, by (25-a), starts with the
\textbf{second} cu\-mu\-lant \hfill \break $Av_c(J^{2}_{(i,j)})=Av(J^{2}_{(i,j)})$,
which is the variance of $J_{(i,j)}$. Condition (25-b) was used to
control the sum in the exponent above.  
 
\textbf{Thermodynamic stability}

Stability considerations play a key role in quantum mechanics
\cite{LiS}. Let a statistical mechanical system be described by a
collection of amiltonians $H_{\Lambda}$, associated to finite regions
$\Lambda \subset \mathbf{R}^{d}$, for particle systems, or $\Lambda
\subset \mathbf{Z}^{d}$ for spin systems, with volume $V(\Lambda)$:
examples are (15) and, for random systems,(24): in the latter case
there is implicit in the Hamiltonians a dependence on the random
variables $J$, and the constant $c$ in definition 2 below is assumed
to be a.e. independent of $J$. The system's free energy is 
$f_{\Lambda}\equiv \frac{-kT\log Z_{\Lambda}(J)}{V(\Lambda)}$ with
$Z_{\Lambda}= tr(\exp(-\beta\ H_{\Lambda}))$, and the thermodynamic
limit means $\Lambda \nearrow \mathbf{R}^{d}$ (or $\mathbf{Z}^{d}$)
with the proviso of fixed density for particle systems, where $\Lambda
\nearrow \mathbf{Z}^{d}$ denotes a limit in the sense of van Hove or
Fisher (the latter being required for random systems, see the
discussion in \cite{vEvH}: (23-d) satisfies this assumption). 

\textbf{Definition 2} 
The system is said to be \textbf{thermodynamically stable} if there
exists a constant  
$0 \le c < \infty$ such that
$$
f_{\Lambda} \ge -c
\eqno{(28-a)}
$$
It is said to satisfy \textbf{thermodynamic stability of the first kind} if
$$
H_{\Lambda} > - \infty
\eqno{(28-b)}
$$
and to satisfy \textbf{thermodynamic stability of the second kind} if
$$
H_{\Lambda} \ge -c V(\Lambda)
\eqno{(28-c)}
$$
The above definitions (28-b,c) are patterned after the corresponding
ones for $N$- body systems in \cite{LiS}: they also apply to
relativistic quantum field theory, where the particle number $N$ is
not conserved. 

Theorem 5 illustrates the interesting fact that, for disordered
systems, stability of the second kind (28-c) may fail even for a
thermodynamically stable system: this happens when only (23-b) (but
not (23-a)) holds. In the following, we shall demonstrate that this
phenomenon has important \textbf{dynamical} implications for a
relevant class of disordered systems. 

We now return to our nonrandom model (15). When (23-a) holds, it
follows from a folklore theorem (see, e.g., Theorem 6-1 of \cite{MWB}
or Theorem 3.3, pg. 111, of \cite{Dav}) and a representation of $f$
(\cite{Ra}, pg.295, and proposition 1) that exponential decay in the
sense that 
$$
|f(t)| \le C \exp(-d|t|) 
\eqno{(29)}
$$
with $C$ and $d$ positive constants, \textbf{cannot} hold. This is
essentially the condition that the physical (GNS) Hamiltonian
$\tilde{H_{0}}$ is bounded from below (semibounded) (see again
\cite{Se3}), where 
$$
\tilde{H_{0}} = \sigma_{0}^{z} \sum_{k \in \mathbf{Z}^{d}} \epsilon(k) \sigma_{k}^{z}
\eqno{(30-a)}
$$
The infinite sum (30) stands for a limit in norm in the quasi-local
algebra ${\cal A}$ associated to the spin system, see  \cite{Ra},
loc. cit. It is the thermodynamic stability of the second kind (28-c)
which guarantees the existence of $\tilde{H_{0}}$, and (29) as a
consequence. In the random case (24), 
$$
\tilde{H_{0}}(J) = \sigma_{0}^{z} \sum_{k \in \mathbf{Z}^{d}} J_{0,k} \epsilon(k) \sigma_{k}^{z}
\eqno{(30-b)}
$$
under assumption (23-a): if only (23-b) holds, the representation
(30-b) is, of course, not defined. We refer to \cite{Wr5}, proposition
4.1 for the dynamical consequences of this fact: in particular, unlike
the stable case (23-a), exponential decay (29) does indeed hold for a
class of one-dimensional potentials with algebraic decay, as shown
there. 

We now consider the special nearest-neighbor interaction (25-c) -
i.e., the EA model of section 3. The following theorem may be proved
(along the same lines, but simpler, than theorem 3.2 of
\cite{Wr5}):\footnote[1]{There are two misprints in the latter
  reference: in the r.h.s. of (40a) the number four should be inside
  the exp function, while in (40b) the first parenthesis should be
  moved to the left of the symbol Av, and the last parenthesis
  omitted.}  

\textbf{Theorem 6} Let (25-c) hold, i.e., (24) describe the EA spin
glass model of section 3. Then there exists a subsequence
$\{m_{n}\}_{n \in \mathbf{Z}}$ such that, a.e. with respect to $J$, 
\begin{eqnarray*}
f(t) \equiv \lim_{m_{n} \to \infty} \rho_{\infty}^{t}(\frac{\sum_{i
    \in V_{n}}\sigma_{i}^{x}}{V_{n}}\\ 
= \delta Av(\wp_{0}(t))
\end{eqnarray*}
$$\eqno{(31)}$$
where
$$
\delta = \phi_{0}(\sigma_{0}^{x}) \ne 0
\eqno{(32)}
$$
and
$$
\wp_{0}(t) \equiv \prod_{\pm\delta_{i}}\cos(2 \beta J_{0,\pm\delta_{i}}t)
\eqno{(33)}
$$
We have the

\textbf{Corollary 6} Consider the distribution functions
$$
dP_{1}(x) = 1/2[\delta_{1}(x) + \delta_{-1}(x)] \mbox{ Bernoulli distribution } 
\eqno{(34-a)}
$$
$$
dP_{2}/dx = 1/2 \chi_{[-1,1]}(x)  \mbox{ uniform distribution } 
\eqno{(34-b)}
$$
$$
dP_{3}/dx = \frac{1}{\sqrt(\pi)} \exp(-x^{2})  \mbox{ Gaussian of unit
  variance } 
\eqno{(34-c)}
$$
The corresponding values of $f$, defined by the r.h.s. of (31), are:
$$
f_{1}(t) = (\cos(2 \beta t)^{z}
\eqno{(35-a)}
$$
$$
f_{2}(t) = (\frac{\sin (2 \beta t)}{2t})^{z}
\eqno{(35-b)}
$$
$$
f_{3}(t) = \exp(-2 z t^{2})
\eqno{(35-c)}
$$

\textbf{Remark 2}

a.) In the Bernoulli case, (35-a) implies no decay, as in the
non-random model: this is an instance of a result of Radin for general
interactions satisfying (23-a), according to which $f$ is
almost-periodic (\cite{Ra}, Lemma 2, pg.1951). 

b.) For the uniform distribution, (35-b) describes a
\textbf{non-Markovian return to equilibrium}. Clearly, in this case,
under assumption (23-a), (30b) is a physical Hamiltonian which is
semibounded for each realization of $J$, and thus exponential decay
(29) cannot hold by the aforementioned folklore theorem. 

c.) In the Gaussian case, not even thermodynamic stability of the
first kind (28-b) holds, because the Gaussian r.v. range over
$\mathbf{R}$. (30b) shows that also the physical Hamiltonian
$\tilde{H_{0}}$ is not bounded below, and thus (29) may, in principle,
hold. That this is actually the case is, of course, made clear by
(35-c). 

d.) Since the transverse magnetization in (31) is an observable (measurable quantity), self-averaging is an important requirement, as discussed before. We conjecture that all subsequential limits are equal, a.e. $J$, to the r.h.s. of (31), i.e., that the limit $n \to \infty$ in (31) exists.

\textbf{The issue of probability distributions}

As is well-known, the use of Gaussian probability distributions (p.d.)
in disordered systems is standard: it simplifies several passages
considerably, e.g., the integration by parts formula in the first
proof of the existence of the thermodynamic limit for the mean field
theory in \cite{GTo}, or in the first proof of diffusion in the full
Anderson model \cite{FS} (the proof here allowed also the uniform
distribution (34-b), considered by Anderson \cite{An1}). One might
think that the results are expected to depend \textbf{qualitatively}
on the probability distribution: this is, however, not so. 

In \cite{SaWr}, where a mean-field Ising model in a random external
field was studied, it was proved that the existence or not of a
tricritical point in the model's phase diagram is tied to the
probability distribution: it is absent for a Gaussian p.d., and
present for a Bernoulli distribution. 

As an attempt to explain this result physically, we conjectured that
it was due to the fact that a discrete distribution of probabilities
such as the Bernoulli distribution samples just a few values of the
couplings and thus introduces some short-ranged elements into the
problem: from this point of view, discrete distributions may have a
closer connection with real materials. This feature is shared by the
uniform distribution, see remark 2 b.), but even here there is a
remarkable \textbf{dynamical} difference between Bernoulli and uniform
distributions with regard to decay, exhibited by (35-a) and
(35-b). The latter yields, somewhat surprisingly, the same result in
one dimension ($z=2$) as obtained by Emch \cite{Em} with an
interaction of infinite range ($\epsilon(|n|) = 2^{-|n|-1}$) in the
nonrandom model (or, alternatively, the random model with Bernoulli
distribution and the same potential). Since the latter (algebraic slow
decay with oscillations) seemed to lead to a good qualitative
description of the nuclear spin resonance experiment \cite{LNo}, the
uniform distribution decay (35-b) might well be in qualitative
agreement with the previously proposed spin glass experiment! 

In contrast to a discrete or uniform distribution, a Gaussian
distribution samples many values of the couplings and works to
reinforce the long-range nature of the interactions. For effectively
short-range inteactions, this might not be suitable.  In any case, the
Gaussian probability distribution leads to a spectacularly fast rate
of return to equilibrium (35-c), and such sharp differences, such as
between (35-c) and (35-b), should be amenable to experiment. 
\bibliography{minhabiblio-t}
\bibliographystyle{alpha}
\end{document}